\documentclass[10pt,twocolumn,twoside]{IEEEtran} 

\usepackage{amsfonts}
\usepackage{amssymb}
\usepackage{amsmath}
\usepackage{amsthm}
\usepackage{graphicx}
\usepackage{subfigure}
\usepackage{multirow}
\usepackage[verbose,nospace,sort]{cite}
\usepackage{algpseudocode}
\usepackage{algorithm}
\usepackage{setspace}

\begin{document}
\theoremstyle{definition}
\newtheorem{theorem}{Theorem}
\theoremstyle{definition}
\newtheorem{example}{Example}
\theoremstyle{definition}
\newtheorem{remark}{Remark}
\newcommand{\xB}{\mathbf{B}}
\newcommand{\xE}{\text{E}}
\newcommand{\xe}{\mathbf{e}}
\newcommand{\xGF}{\text{GF}}
\newcommand{\xtr}{\text{tr}}
\newcommand{\xd}{\mathbf{d}}
\newcommand{\xH}{\mathbf{H}}
\newcommand{\xI}{\mathbf{I}}
\newcommand{\xtI}{\text{I}}
\newcommand{\xtlog}{\text{log}}
\newcommand{\xmkl}{_{k,l}}
\newcommand{\xmkj}{_{k,j}}
\newcommand{\xQ}{\mathbf{Q}}
\newcommand{\xP}{\mathbf{P}}
\newcommand{\xp}{\mathbf{p}}
\newcommand{\xR}{\mathbf{R}}
\newcommand{\xSF}{\text{SF}}
\newcommand{\xT}{\mathbf{T}}
\newcommand{\xtSINR}{\text{SINR}}
\newcommand{\xu}{\mathbf{u}}
\newcommand{\xU}{\mathbf{U}}
\newcommand{\xv}{\mathbf{v}}
\newcommand{\xtv}{\text{v}}
\newcommand{\xV}{\mathbf{V}}
\newcommand{\xw}{\mathbf{w}}
\newcommand{\xy}{\mathbf{y}}
\newcommand{\xX}{\mathbf{X}}
\newcommand{\xx}{\mathbf{x}}
\newcommand{\xz}{\mathbf{z}}

\title{Balancing Weighted Substreams in MIMO Interference Channels}

\author{Cenk M. Yetis, \IEEEmembership{Member, IEEE,
} Yong Zeng, \IEEEmembership{Student Member, IEEE, } Kushal Anand, \IEEEmembership{Student Member, IEEE, }
\\ Yong Liang Guan, \IEEEmembership{Member, IEEE} and Erry Gunawan, \IEEEmembership{Member, IEEE}
\thanks{The work of C. M. Yetis in part and Y. Zeng, K. Anand, Y. L. Guan, and E. Gunawan in full was supported by the Department of the Navy Grant N62909-12-1-7015 issued by Office of Naval Research Global.}
\thanks{C. M. Yetis is with the Department of Electrical and Electronics Engineering, Mevlana University, Turkey (e-mail:cenkmyetis@ieee.org). Further documents pertinent this letter are available on the author's website http://sites.google.com/site/cenkmyetis.}
\thanks{Y. Zeng is with the Department of Electrical and Computer Engineering, National University of Singapore (e-mail:elezeng@nus.edu.sg).}
\thanks{K. Anand, Y. L. Guan, and E. Gunawan are with the School of Electrical and Electronic Engineering, Nanyang Technological University, Singapore (e-mail:\{kush0005\}@e.ntu.edu.sg, \{eylguan, egunawan\}@ntu.edu.sg).}}
\maketitle
\begin{abstract}
Substreams refer to the streams of each user in a system. Substream weighting, where the weights determine the prioritization
order, can be important in multiple-input multiple-output interference channels. In this letter, a distributed algorithm is
proposed for the problem of power minimization subject to weighted SINR constraint. The algorithm is based on two basic
features, the well known distributed power control algorithm by Yates in 1995 and a simple linear search to find feasible SINR
targets. The power control law used in the proposed algorithm is proven to linearly converge to a unique fixed-point.
\end{abstract}
\begin{keywords}
MIMO, interference channel, weighted SINR, substream balancing.
\end{keywords}

\section{Introduction} \label{sec:Intro}
Prioritization, in other words desired level of fairness, is important to ensure quality-of-service (QoS) in the
system. Balancing weighted substreams, data streams of the same user, can be achieved by two complementary
approaches, maximization of minimum weighted SINR subject to power constraint or minimization of power subject
to weighted SINR constraint, and at three different levels, prioritization between streams, users, or
substreams. Explicitly, prioritization between all streams, all users, or all streams of each user can be aimed.
From more to less restrictive, stream, user, and substream prioritization comes in order. Consequently,
\mbox{substream} prioritization causes the least degradation in \mbox{sum-rate}, followed by user and stream
prioritization for a given channel. In this work, \mbox{substream} prioritization is studied.

Unequal and equal weighting of SINRs are both important in practice. The former aims for the received SINRs of more important
substreams to be higher than those of less important substreams. On the other hand, the later aims for a more error resilient
system. If a substream cannot be decoded, e.g., the information is lost, other substreams can be used to achieve a successful
transmission with a lower quality. To this end, in this letter, a distributed algorithm is proposed to achieve desired level
of SINR fairness at the substream level by using the later approach, minimizing the power subject to weighted SINR constraint.
The proposed algorithm is \mbox{ad-hoc} in the sense that the transmit and receive beamforming vectors are initially obtained
via a beamforming scheme such as SINR maximization (\mbox{max-SINR}) \cite{55}. Then, the proposed algorithm is plugged and
run in an \mbox{ad-hoc} manner. The proposed algorithm has two basic features. The power control law, the first feature, used
in the algorithm is a straightforward extension of standard interference functions introduced in \cite{207} as was also
applied in \cite{333,224}. The linear search, the second feature, used in the algorithm finds feasible SINR targets for the
substreams, thus convergence of the algorithm is guaranteed. The contributions of the letter can be summarized as follows. 1)
The ad-hoc nature of the proposed algorithm allows linear search for setting the SINR targets dynamically, as opposed to
setting the SINR targets statically. To the best of our knowledge, setting SINR targets opportunistically in multiple-input
multiple-output (MIMO) interference channels (ICs) is only studied in this letter. 2) A system with unequal substream priority
can transmit a media with higher quality than a system with equal substream priority although total bit error rate (BER) of
the first approach can be higher than the second. On the opposite, total BER is influential for the received media quality
under equal substream priority, in other words under substream fairness condition \cite{220}. For the later, this letter takes
initiative steps in MIMO ICs in showing the effects of substream fairness on uncoded BER, SINR and \mbox{sum-rate} metrics. 3)
Power control law in the proposed algorithm based on standard interference functions is proven to converge to a unique
fixed-point. In fact, the convergence is independent from beamforming techniques and desired level of fairness.

Notation: $^\text{T}$ and $^\dagger$ denote the transpose and complex conjugate operators. Matrices and vectors
are denoted by bold-face uppercase and lowercase letters, respectively. $\mathbf{1}$, $\xI$, $\mathbf{0}$, and
diag$[x_1,\ldots,x_l,\ldots, x_L]$ denotes all ones vector, identity matrix,  zero vector or matrix, and
diagonal matrix with elements $x_l$ on its diagonal, respectively. $|.|$, $||.||_1$, and min denote determinant,
$l_1$-norm, and minimum operators, respectively, and for some given vector $\textbf{x}>\mathbf{0}$,
$||.||_\infty^{\text{x}}$ denotes weighted maximum norm.

\section{System Model} \label{sec:Model}

We consider a $K$-user IC, where there are $K$ transmitters and receivers with $M_k$ and $N_k$ antennas at node
$k$, respectively. A transmitter has $d_k$ streams to be sent to its corresponding receiver. This system can be
modeled as \mbox{$\xy_k=\sum_{j=1}^K\xH_{kj}\xx_j+\xz_k$},$~ \forall k\in\mathcal{K}\triangleq\{1,2,...,K\},$
where $\xy_k \textrm{ and } \xz_k$ are the $N_k\times 1$ received signal vector and the zero mean unit variance
circularly symmetric additive white Gaussian noise vector (AWGN) at the $k^{th}$ receiver, respectively. $\xx_j$
is the $M_j\times 1$ signal vector transmitted from the $j^{th}$ transmitter and $\xH_{kj}$ is the $N_k\times
M_j$ channel matrix between the $j^{th}$ transmitter and the $k^{th}$ receiver. $\xE[||\xx_j||^2]=p_j$ is the
power of the $j^{th}$ transmitter. The transmitted signal from the $j^{th}$ user is
$\xx_j=\xU_j\sqrt{\xP_j}\xd_j$, where $\xU_j=[\xu_{j,1},\ldots,\xu_{j,d_j}]$ is the $M_j\times d_j$ precoding
matrix, $\xd_j$ is $d_j\times 1$ vector denoting the $d_j$ independently encoded streams, and
$\xP_j=\text{diag}[p_{j,1},\ldots,p_{j,d_j}]$ is a $d_j\times d_j$ diagonal matrix consisting of substream
powers with $\sum_{l=1}^{d_j}p_{j,l}\leq p_j$. The $N_k\times d_k$ receiver matrix is denoted by $\xV_k$.

\section{Prioritized Substreams} \label{sec:Prioritization}
Substream prioritization is useful for concurrent transmissions of different services, e.g., voice and media
transmissions, as well as transmission of single service, e.g., different parts of the video or image can be
assigned to substreams with varying importance. On the other hand, all substreams can have equal importance to
have error resiliency. For both, the proposed \mbox{ad-hoc} algorithm in this letter can dynamically set the
SINR targets. Thus, the algorithm has small rate and SINR losses.

\subsection{Preliminaries} \label{subsec:Preliminaries}
\mbox{Max-SINR} \cite{55} designs the transmit beamformer of each stream of a user separately, substreams are considered as
interference on one another, thus SINR is given as
$\xtSINR\xmkl=\frac{\xv\xmkl^\dagger\xR\xmkl\xv\xmkl}{\xv\xmkl^\dagger\xB\xmkl\xv\xmkl}$, where $\xB\xmkl=\xQ\xmkl+\xI_{N_k}$,
$\xQ\xmkl=\sum_{j=1}^K\xH_{kj}\xU_j\xP_j\xU_j^\dagger\xH_{kj}^\dagger-\xR\xmkl$, and
$\xR\xmkl=p\xmkl\xH_{kk}\xu\xmkl\xu\xmkl^\dagger\xH_{kk}^\dagger$ are the covariance matrices of the interference plus noise,
interference, and $l^{th}$ stream of user $k$, respectively.

A simple technique for substream prioritization is weighting the \mbox{substream-SINRs}, thus substream prioritization can be
achieved via joint power control and beamforming design by maximizing the minimum weighted SINRs \cite{327}. Compared with
stream prioritization problems, the substream problems are decoupled into $K$ \mbox{sub-problems} given SINR targets are
feasible, then the problem can be solved asynchronously among users. However, feasibility check is coupled among users and can
be shown to be \mbox{NP-hard} \cite{327}. Therefore, we focus on designing efficient algorithms for achieving locally optimal
points. It is well known that the optimal solution to the minimization of power subject to weighted SINR constraint is
achieved when the weighted SINRs are equalized, i.e.,
$\frac{\xtSINR_{k,1}}{\beta_{k,1}}=\cdots=\frac{\xtSINR_{k,d_k}}{\beta_{k,d_k}}=\Gamma_{k}^\text{C},~\forall k\in\mathcal{K}$,
where $\beta_{k,l}$ are the weighting factors that reflect the priorities and $\Gamma_{k}^\text{C}$ is the common weighted
SINR target of substreams. Clearly, when $\beta_{k,l}=1, \forall k\in\mathcal{K}\text{ and } \forall
l\in\mathcal{L}\triangleq\{1,2,...,d_k\}$, the problem is reduced to conventional worst SINR maximization problem. We propose
a practical scheme, named \mbox{ad-hoc} algorithm, to balance weighted \mbox{substream-SINRs}. Basically, we unite the simple
linear search for finding maximum possible SINR targets with the optimization problem
\begin{align*}
    &\min\sum_{l=1}^{d_k}p\xmkl \text{ subject to }\\& \frac{\xtSINR\xmkl}{\beta\xmkl}\geq \Gamma_{k}^\text{C}, p\xmkl>0, \sum_{l=1}^{d_k}p\xmkl\leq p_k, \forall k\in\mathcal{K} \text{ and } \forall l\in\mathcal{L}
\end{align*} 
that can be solved via conventional distributed power control algorithm \cite{207}\cite{333}\cite{224} with the maximum power
constraint. The well known distributed power control algorithm with maximum power per user $p_k$ constraint \cite{207} is
given as $p_k^n=\text{min}\left(\frac{\Gamma_k}{\xtSINR_k^{n-1}}p_k^{n-1},p_k\right),$ where superscript $n$ is the iteration
number, $p_k^n$ is the power, and $\xtSINR_k^{n-1}$ is the SINR of user $k$. Basically, a user increases its power if its SINR
is below its SINR target and vice versa. Clearly the SINR target can be unmet due to the maximum power constraint. The goal of
the proposed algorithm is to achieve substream prioritization while causing the least \mbox{sum-rate} degradation. Therefore,
power saving is not the primary concern of our proposed algorithm. By directly following the steps in \cite{207}, the standard
interference function for our problem is given as \cite{333,224}
\begin{equation}\label{eqn:ourSIF}
\xtI\xmkl(\xp)=\Gamma\xmkl\delta\xmkl,
\end{equation}
where $\xp=[p_{1,1},\ldots,p_{1,d_1},\ldots,p_{K,1},\ldots,p_{K,d_K}]^\text{T}$ is the transmitted power vector
of the system, $\Gamma\xmkl=\beta\xmkl\Gamma_k^\text{C}$, and $\delta\xmkl=\frac{p\xmkl}{\xtSINR\xmkl}.$

Finally, joint optimization of beamforming vectors and power allocation is a challenging problem for schemes
where SINR targets are dynamically determined. The degrees of freedom (DoF) essentially drop to zero for schemes
where SINR targets are preset \cite{327,224}. Whereas DoF is not zero for our scheme since opportunistic maximum
SINR search is performed, together with achieving fairness between data streams. This is shown in
\mbox{sum-rate} simulation results in Section \ref{sec:NumericalResults} where the DoF loss is not significant
compared to the conventional \mbox{max-SINR} where fairness is not achieved.

\subsection{Proposed Algorithm}
The proposed \mbox{ad-hoc} algorithm in Table \ref{alg:DistAlg} opportunistically searches for feasible SINR targets for
substreams. The algorithm can run asynchronously among users. In Table \ref{alg:DistAlg},
$\xp_k^n=[p_{k,1}^n,\ldots,p_{k,d_k}^n]$ is the power vector, $p\xmkl^n$ is the power at iteration $n$,
$\beta_{k,l}^\text{N}=\beta_{k,l}/\sum_{l=1}^{d_k}\beta_{k,l}$ is the normalized weighting factor for the $l^{th}$ substream
of the $k^{th}$ user, and $\epsilon$ is set to $10^{-3}$. $\xP_j^{n-1}$ and $\xB^{n-1}\xmkl$ are the previously defined terms
with the iteration numbers. $\xR\xmkl^\prime=\xH_{kk}\xu\xmkl\xu\xmkl^\dagger\xH_{kk}^\dagger$ is akin to a covariance matrix,
$\mathbf{1}=[1,\ldots,1]^\text{T}$ is all ones vector, $\boldsymbol{\delta}_k\triangleq[\delta_{k,1},\ldots,\delta_{k,d_k}]$,
and similarly $\textbf{SINR}_k$ is the vector of \mbox{substream-SINRs} for user $k$. $p_{k,\text{total}}$ is a variable used
in the simulation that can have maximum value $p_k$, and the limit variable is an upper bound for the iteration number of
power control law.

\begin{table}[htb!]\footnotesize
\caption{Distributed Ad-Hoc Algorithm} \label{alg:DistAlg} 
\begin{algorithmic}[1]
\State Evaluate SINR outcomes of max-SINR beamforming scheme, $\text{SINR}_k'$ \State $\%~m=0$, initialize
$\text{SINR}_{k}=\text{SINR}_k'$, $\forall k\in \mathcal{K} $, check=0 \While {check$\sim$=1}\State $\%~m=m+1$,
$\xp_k^0=\frac{p_k}{d_k}\mathbf{1}$, $\xp_k^1=2\xp_k^0$, $n=1$ \State
$\Gamma_{k,l}=\beta_{k,l}^\text{N}\text{SINR}_k$, $\forall k\in \mathcal{K}$, $\forall
l\in\mathcal{L}$\While{$\sum_{k=1}^K||\xp_k^n-\xp_k^{n-1}||_1>\epsilon$ \& $n\leq \text{limit}$ } \State
$\delta\xmkl=\frac{\xv\xmkl^\dagger\xB^{n-1}\xmkl\xv\xmkl}{\xv\xmkl^\dagger\xR\xmkl'\xv\xmkl}$, $\forall k\in
\mathcal{K}, \forall l\in\mathcal{L} $ \State $x=2\max(\boldsymbol{\delta}_k)$, $p_{k,\text{total}}=0$, $\forall
k\in \mathcal{K}$ \For {counter=1:$d_k$}, $\forall k\in \mathcal{K}$ \State
$[\sim,y]=\min(\boldsymbol{\delta}_k)$ \label{step:min} \State
$p_{k,y}^n=\min(\Gamma_{k,y}\delta_{k,y},p_k-p_{k,\text{total}})$ \State
$p_{k,\text{total}}=p_{k,\text{total}}+p_{k,y}^n$, $\delta_{k,y}=x$ \EndFor, $n=n+1$\EndWhile \State Evaluate
new SINRs $\text{SINR}\xmkl$ by using new power values $\xp_k^n$, $\forall k\in \mathcal{K}$ \State
$\Delta_k=(\sum_{l=1}^{d_k}\text{SINR}_{k,l}/\beta_{k,l})/d_k-\text{min}(\textbf{SINR}_k)$, $\forall k\in
\mathcal{K}$ \If {$\sum_{k=1}^K\Delta_k\leq\epsilon$} check=1 \EndIf \EndWhile
\end{algorithmic}
\end{table}

\subsubsection*{Linear search}
Since the proposed algorithm is ad-hoc, the maximum SINR achieved after beamforming is the upper bound to the maximum SINR
achieved after the proposed \mbox{ad-hoc} algorithm. For simplicity, consider the substream fairness constraint case, where
$\beta_{k,l}=1,~\forall l\in\{1,2\}$. Thus $\Gamma_k^\text{C}=\overline{\Gamma}_k$, where $\overline{\Gamma}_k$ is average
SINR of user $k$. Clearly setting average SINR as the target is a good starting point for searching as will be shown in
Example \ref{ex:LinSearch}.

\example \label{ex:AvSINR} Assume $\xtSINR_{k,1}^0=5$ and $\xtSINR_{k,2}^0=10$ are achieved for the $k^{th}$
user after beamforming. Thus for \mbox{substream} fairness, SINR target is set to
$\Gamma\xmkl^1=\overline{\Gamma}_k^1=7.5,~\forall l\in\{1,2\}$, where superscript denotes the iteration number
$m=1$, and achieved SINR after the first iteration is denoted by $\xtSINR_{k,l}^1$. Please note that iteration
number $m$ is not needed in Algorithm \ref{alg:DistAlg} thus it is omitted. Ideally
$\xtSINR_{k,1}^m=\xtSINR_{k,2}^m=7.5$ must be achieved after the \mbox{ad-hoc} algorithm. However, due to the
distributed nature of the algorithm, and especially when the substreams are highly unbalanced, the optimal SINR
target may not be achieved.

Step \ref{step:min} of the algorithm is the most critical part where the substream powers are updated in order
from the substream with the lowest to the highest $\delta\xmkl$. In this way, the substream with the lowest
$\delta\xmkl$ can definitely reach the SINR target, while the substream with the highest $\delta\xmkl$ reaches
to a maximum possible SINR value by using the remaining power budget of the user $k$. In the next iteration, the
target SINR is the average of these achieved SINRs, thus the algorithm keeps iterating until the convergence of
\mbox{substream-SINRs}. The convergence plot of the proposed algorithm is given in \cite{220}.

\example \label{ex:LinSearch} As explained previously, the substream with the highest SINR is guaranteed to
achieve the SINR target $\overline{\Gamma}_k^m$ in the $m^{th}$ iteration. Consider Example \ref{ex:AvSINR}.
After the first iteration, the second substream is guaranteed to achieve the target
$\xtSINR_{k,2}^1=\overline{\Gamma}_k^1=7.5$. Meanwhile, assume that only $\xtSINR_{k,1}^1=5.5$ can be achieved
for the first substream. Then the SINR target for the next iteration is $\overline{\Gamma}_k^2=6.5$. After the
second iteration, $\xtSINR_{k,2}^2=6.5$ is again guaranteed to be achieved and assume only
$\xtSINR_{k,1}^2=5.75$ can be achieved by expending the remaining power. Then for the third iteration, SINR
target is $\overline{\Gamma}_k^3=6.125$. The SINR target keeps dropping until the \mbox{substream-SINRs} are
equal, thus convergence is guaranteed.

Finally, in the high SNR regime, e.g., at 30 dB, high number of iterations for power control law, i.e., while
loop  between the lines 6 and 14 in Algorithm \ref{alg:DistAlg}, can be required. An upper bound can be set and
the parameter $\epsilon$ can be tuned for avoiding high number of iterations. For further details, the reader is
referred to the first author's website \cite{220}.

\section{Numerical Results} \label{sec:NumericalResults}

Numerical results for MIMO ICs with $K=3, M_k=4, N_k=4,\text{ and } d_k=2$ are presented in this section. $10^3, 10^4, 10^5,
\text{ and } 10^6$ random ICs are tested for SNR values 0, 5, 10, and 15 dB, respectively. Channel coefficients are generated
by i.i.d. zero-mean \mbox{unit-variance} complex Gaussian variables, QPSK modulation is used, and iteration number is set to
16. The \mbox{sum-rate} is defined as $\text{R}_{\text{sum}}=\sum_{k=1}^{K}\sum_{l=1}^{d_k}\log_2(1+\xtSINR\xmkl)$ \cite{204}.
Finally, the same type of filter structures are used at the transmitters and receivers, i.e., \mbox{max-SINR} filter is used
for both transmit and receive filters.

\subsection{Balancing weighted substreams}

In Table \ref{tabl:BWSubstreams}, for $\beta_{k,1}=1$ and $\beta_{k,2}=6,~ \forall k\in\mathcal{K}$, SINR values
and ratios before ($\text{SINR}_{k,l}^0$) and after ($\text{SINR}_{k,l}^m$) Algorithm \ref{alg:DistAlg} are
presented at 10 dB. Please note that given predetermined priorities $\beta_{k,l}$, \textit{optimal} balancing of
weighted substreams can be achieved if the \mbox{substream-SINR} proportions are already smaller than the
priority proportions, i.e.,
\mbox{$\frac{\text{SINR}_{k,2}^0}{\text{SINR}_{k,1}^0}<\frac{\beta_{k,2}}{\beta_{k,1}}$}.

\begin{table} [htb!]
  \begin{center} \caption{Balancing weighted substreams.} \label{tabl:BWSubstreams}
\begin{tabular}{|c|c|c|c|c|c|c|c|}
  \hline
   User $k$ & $\text{SINR}_{k,1}^0$ & $\text{SINR}_{k,2}^0$ & $\text{Ratio}^0$ & $\text{SINR}_{k,1}^m$ & $\text{SINR}_{k,2}^m$  & $\text{Ratio}^m$ \\
     \hline
1 & 8.89 & 22.53 & 2.53 & 4.49 & 26.93 & 6 \\
     \hline
2 & 8.09 & 21.54 & 2.66 & 4.23 & 25.40 & 6 \\
     \hline
3 & 8.13 & 19.49 & 2.40 & 3.95 & 23.67 & 6 \\
     \hline
\end{tabular}
    \end{center}
\end{table}

\subsection{Achieving substream fairness}
\begin{figure}[htb]
\centering
    \subfigure[Stream-BER comparison.] {
\includegraphics[height=4.25cm, width=8cm] {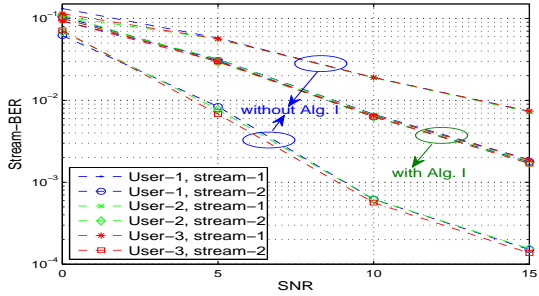} \label{Fig:StreamBERs}
}
   \subfigure[BER comparison.] {
\includegraphics[height=4.25cm, width=8cm] {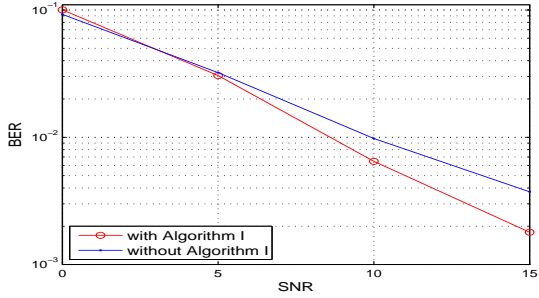} \label{Fig:BER}
} \caption{Bit error rates of max-SINR. } \label{Fig:Errorrates}
\end{figure}

\begin{figure}[htb]
\centering
\includegraphics[height=4.25cm, width=8cm] {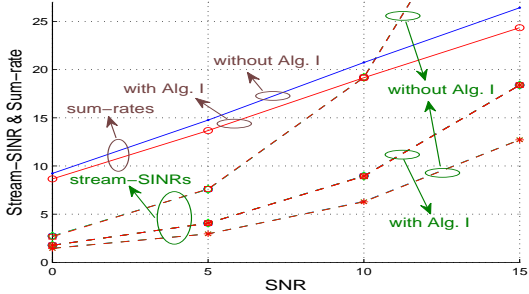}
 \caption{Stream-SINRs and sum-rates of max-SINR.} \label{Fig:SINRsRates}
\end{figure}

For \mbox{max-SINR} with and without Algorithm \ref{alg:DistAlg}, uncoded BER results in Fig. \ref{Fig:Errorrates}, and
\mbox{stream-SINRs} and \mbox{sum-rates} concurrently in Fig. \ref{Fig:SINRsRates} are plotted where weighting factors are set
to $\beta_{k,l}=1, ~ \forall k\in\mathcal{K} \text{ and } \forall l\in\mathcal{L}$. For \mbox{max-SINR} without Algorithm
\ref{alg:DistAlg}, power is simply split equally among the substreams of each user. SINR and \mbox{sum-rate} results, and SNR
values are in linear scale, bits per channel use, and in dB scale, respectively. SINR legends in Fig. \ref{Fig:SINRsRates}
follow the same order in Fig. \ref{Fig:StreamBERs}. The second \mbox{substream-SINRs} of \mbox{max-SINR} without Algorithm
\ref{alg:DistAlg} achieve values around 45 at 15 dB, but not plotted for brevity. Substream fairness is achieved at the cost
of a reasonable \mbox{sum-rate} degradation as seen in Fig. \ref{Fig:SINRsRates}. The proposed algorithm whose objective is
substream fairness can achieve stream fairness in ergodic sense as seen in Fig. \ref{Fig:SINRsRates}, with lower algorithmic
complexity and less information exchange than the algorithms whose objectives are stream fairness.

%

\section{Rate of Convergence} \label{sec:ROC}
In this section, the power control law used in Algorithm \ref{alg:DistAlg} is proven to converge to a unique
fixed-point at a linear rate under appropriate assumptions by using contractive interference functions
introduced in \cite{225}. For further details regarding this section, the reader is referred to \cite{225}.
Next, necessary steps are taken to prove interference functions \eqref{eqn:ourSIF} are contractive.

The interference function \eqref{eqn:ourSIF} can be rewritten as $\xtI\xmkl(\xp)=\sum_{s=1}^{d_j}\sum_{j=1}^{K}T_{k,l}^{j,s}
p_{j,s}+N_{k,l}$, where $N_{k,l}=\frac{\Gamma_{k,l}}{G_{k,l}^{k,l}},$ $G_{k,l}^{j,s}=|\xv\xmkl^\dagger\xH\xmkj\xu_{j,s}|^2$,
and
\begin{equation}\label{eqn:entitites}
 T_{k,l}^{j,s} =
  \begin{cases}
0       & \text{if } (k,l)=(j,s),
    \\
\frac{\Gamma_{k,l} G_{k,l}^{j,s}}{G_{k,l}^{k,l}} & \text{else}.
  \end{cases}
\end{equation}
Define $\xT\left(\sum_{m=1}^{k-1}d_m+l,\sum_{n=1}^{j-1}d_n+s\right)=T_{k,l}^{j,s}$ as a \mbox{$Kd\times Kd$}
matrix with entities in \eqref{eqn:entitites}, where $a$ and $b$ in $\xT(a,b)$ denote the row and column
indices, respectively \cite{333}.

\theorem If $||\xT||_\infty^{\xtv}<1$ for some $\textbf{v}>\textbf{0}$, then interference functions \eqref{eqn:ourSIF} are
\textit{c}-contractive interference functions with \mbox{$c=||\xT||_\infty^{\xtv}$}.
\begin{proof}
The interference functions satisfy the contractivity condition with \mbox{$c=||\xT||_\infty^{\xtv}$}
\begin{eqnarray*}
  \xtI\xmkl(\xp+\epsilon\xv) &=& \xtI\xmkl(\xp)+\epsilon\sum_{s=1}^{d_j}\sum_{j=1}^{K}T_{k,l}^{j,s}v_j \\
   &\leq& \xtI\xmkl(\xp)+\epsilon||\xT||_\infty^{\xtv}
v_k.
\end{eqnarray*}
\end{proof}
An easily verifiable but a more conservative choice for $\xv$ can be the $\xv=1$ option. In this case, the row
sums or the spectral radius of the matrix $\xT$ should be less than 1 \cite{225}. We refer the interested reader
to the first author's website \cite{220} for a more comprehensive treatment on the convergence of power control
law.

\section{Conclusion}
A distributed \mbox{ad-hoc} algorithm that balances weighted \mbox{substream-SINRs} has been developed. The
algorithm guarantees feasible SINR targets opportunistically via its \mbox{ad-hoc} and linear search features.
Via contractive interference functions, the power control law in the proposed algorithm is proven to linearly
converge to a unique fixed-point under appropriate assumptions.

\bibliographystyle{IEEEtran}
\bibliography{IEEEabrv,IEEEfull}

\begin{thebibliography}{1}
\providecommand{\url}[1]{#1}
\csname url@samestyle\endcsname
\providecommand{\newblock}{\relax}
\providecommand{\bibinfo}[2]{#2}
\providecommand{\BIBentrySTDinterwordspacing}{\spaceskip=0pt\relax}
\providecommand{\BIBentryALTinterwordstretchfactor}{4}
\providecommand{\BIBentryALTinterwordspacing}{\spaceskip=\fontdimen2\font plus
\BIBentryALTinterwordstretchfactor\fontdimen3\font minus
  \fontdimen4\font\relax}
\providecommand{\BIBforeignlanguage}[2]{{%
\expandafter\ifx\csname l@#1\endcsname\relax
\typeout{** WARNING: IEEEtran.bst: No hyphenation pattern has been}%
\typeout{** loaded for the language `#1'. Using the pattern for}%
\typeout{** the default language instead.}%
\else
\language=\csname l@#1\endcsname
\fi
#2}}
\providecommand{\BIBdecl}{\relax}
\BIBdecl

\bibitem{55}
K.~S. Gomadam, V.~R. Cadambe, and S.~A. Jafar, ``A distributed numerical
  approach to interference alignment and applications to wireless interference
  networks,'' \emph{{IEEE} Trans. Inf. Theory}, vol.~57, no.~6, pp. 3309--3322,
  June 2011.

\bibitem{207}
R.~D. Yates, ``A framework for uplink power cellular radio systems,''
  \emph{{IEEE} J. Sel. Areas Commun.}, vol.~13, no.~17, pp. 1341--1347, Sep.
  1995.

\bibitem{333}
C.~Wilson and V.~V. Veeravalli, ``A convergent version of the max {SINR}
  algorithm for the {MIMO} interference channel,'' \emph{{IEEE} Trans. Wireless
  Commun.}, vol.~12, no.~6, pp. 2952--2961, Jun. 2013.

\bibitem{224}
H.~Farhadi, C.~Wang, and M.~Skoglund, ``Distributed interference alignment and
  power control for wireless {MIMO} interference networks,'' \emph{{IEEE}
  Wireless Commun. Netw. Conf.}, pp. 3077 -- 3082, 7-10 April 2013.

\bibitem{220}
C.~M. Yetis, Y.~Zeng, K.~Anand, Y.~L. Guan, and E.~Gunawan, ``Sub-stream
  fairness and numerical correctness in {MIMO} interference channels,''
  \emph{{IEEE} Symposium on Wireless Technology and Applications ({ISWTA})},
  pp. 91--96, 22-25 Sept. 2013, http://sites.google.com/site/cenkmyetis/.

\bibitem{327}
Y.~F. Liu, Y.~H. Dai, and Z.~Q. Luo, ``Max-min fairness linear transceiver
  design for a multi-user {MIMO} interference channel,'' \emph{{IEEE} Trans.
  Signal Process.}, vol.~61, no.~9, pp. 2413--2423, May 2013.

\bibitem{204}
H.~Shen, B.~Li, M.~Tao, and X.~Wang, ``{MSE-B}ased transceiver designs for the
  {MIMO} interference channel,'' \emph{{IEEE} Trans. Wireless Commun.}, vol.~9,
  no.~11, pp. 3480--3489, Nov. 2010.

\bibitem{225}
H.~R. Feyzmahdavian, M.~Johansson, and T.~Charalambous, ``Contractive
  interference functions and rates of convergence of distributed power control
  laws,'' \emph{{IEEE} Trans. Wireless Commun.}, vol.~11, no.~12, pp.
  4494--4502, Dec. 2012.

\end{thebibliography}

\end{document}